# Reactive Programming without Functions

Bjarno Oeyen[a] , Joeri De Koster[a] , and Wolfgang De Meuter[a]

a   Software Languages Lab, Vrije Universiteit Brussel, Belgium

**Abstract**

**Context**  Reactive programming (RP) is a declarative programming paradigm suitable for expressing the handling of events. It enables programmers to create applications that react automatically to changes over time. Whenever a time-varying *signal* changes — e.g. in response to values produced by event stream (e.g., sensor data, user input…) — the program state is updated automatically in tandem with that change. This makes RP well-suited for building interactive applications and reactive (soft real-time) systems.

**Inquiry**  RP Language implementations are often built on top of an existing (host) language as an Embedded Domain Specific Language (EDSL). This results in application code in which reactive code and non-reactive code is inherently entangled. Using a mechanism known as *lifting*, one usually has access to the full feature set of the (non-reactive) host language in the RP program. However, lifting is also dangerous. First, host code expressed in a Turing-complete language may diverge, resulting in unresponsive programs: i.e. reactive programs that are not actually reactive. Second, the bi-directional integration of reactive and non-reactive code results in a paradigmatic mismatch that, when unchecked, leads to faulty behaviour in programs.

**Approach**  We propose a new reactive programming language, that has been meticulously designed to be reactive-only. We start with a simple (first-order) model for reactivity, based on *reactors* (i.e. uninstantiated descriptions of signals and their dependencies) and *deployments* (i.e. instances of reactors) that consist of *signals*. The language does not have the notion of functions, and thus unlike other RP languages there is no lifting either. We extend this simple model incrementally with additional features found in other programming languages, RP or otherwise. These features include stateful reactors (that allow for time-based accumulation), signals with dynamic dependencies by means of conditionals and polymorphic deployments, recursively-defined reactors, and (anonymous) reactors with lexical scope.

**Knowledge**  In our description of these language features, we not only describe the syntax and semantics, but also how each features compares to the problems that exist in (EDSL) RP languages. I.e. by starting from a reactive-only model, we identify which reactive features (that, in other RP languages are typically expressed in non-reactive code) affect the *reactive guarantees* that can be enforced by the language.

**Grounding**  We base our arguments by analysing the effect that each feature has on our language: e.g., by analysing how signals are updated, how they are created and how dependencies between signals can be affected. When applicable, we draw parallels with other languages: i.e. similarities shared with other RP languages will be highlighted and thoroughly analysed, and where relevant the same will also be done with non-reactive languages.

**Importance**  Our language shows how a purely reactive programming is able to express the same kinds of programs as in other RP languages that require the use of (unchecked) functions. By considering reactive programs as a collection of pure (reactive-only) reactors, we aim to increase how reactive programming is comprehended by both language designers and its users.

## The Art, Science, and Engineering of Programming

 **Introduction**

Reactive Programming (RP) languages are languages where developers specify *what* time-varying values make up their program, without specifying exactly *how* they are changed over time [3]. RP languages provide a convenient way for variables to be automatically updated if the variables they depend on change (i.e. by re-evaluating the expressions that define them). Reactive programming languages typically call these variables *time-varying signals*, or *signals* for short [15]. The ideas of RP provide an alternative approach to update the various components that make up a large program. Using RP yields distinct advantages over classical event handling approaches like callbacks, which are known to be unsafe to compose [30, 33] as they typically rely on (global) variables that may be read and written to by different callbacks. Correctly using callbacks therefore requires programmers to have a rigorous understanding of the evaluation order of these callbacks, lest they introduce bugs in the event handling code which is a frequent issue in application development [39]. Furthermore, it has been shown that RP programs are easier to write and comprehend than programs with callbacks [45].

In the last decades, many reactive programming languages (and libraries) have emerged, targeting various application domains such as animation [15], GUIs [8, 12], game development [37], robotics [23, 46], networking [31, 55], stage lighting [49] and distributed systems [14, 35, 43, 48]. Many of these languages are built as an embedded domain-specific language (EDSL), i.e. as languages that extend an existing general-purpose language. This approach allows existing code to be integrated with reactive code, and allows programmers to use the ecosystem of the host language (libraries, compilers, IDEs…) [8, 44]. We will refer to these languages as *two-layered* RP languages. The first layer, *the base layer*, governs the non-reactive semantics. This layer corresponds to the host language, which is usually a functional or imperative language. The second layer, *the reactive layer*, governs the reactive semantics. This layer provides the reactive abstractions and the mechanisms to keep them updated.

Next to EDSL implementations, there exist also self-contained RP languages [12, 46, 57]. While these languages do not benefit from an existing host language (e.g., its ecosystem), the language implementation has *complete* control over the semantics of the language. Nonetheless, existing self-contained RP languages are very similar to their EDSL siblings as they are also two-layered: i.e. the reactive language is embedded in a non-reactive subset language. The full language (i.e. the reactive superset language) then uses this subset language to implement the behaviour of the various types of signals that make up programs.

## 1.1 Problem Statement

A central concept in every two-layered language is the notion of *lifting* [15]. Lifting allows programmers to write (base-layer) functions, and to apply them — in lifted form — to signals. When the value of one or more signals changes its value, all lifted functions that have been applied to these signals will be re-executed. As a consequence, EDSL RP languages constantly needs to switch between reacting (i.e. propagating signal updates) and computing (i.e. evaluating lifted code to determine a signal's updated value). This combination of evaluating reactive and non-reactive (lifted) code can be cumbersome and exhibit undesirable behaviour. Moreover, it makes these languages more difficult to formalise and reason about, given the bi-directional semantic dependencies between both layers. These issues have been identified in earlier work [51, 53].

**Reactive Thread Hijacking Problem**  When a two-layered RP language evaluates a lifted function, the reactive runtime needs to wait until the function call returns. Lifted code therefore can hijack the evaluation thread of the reactive runtime (e.g., when the lifted code — implemented in the base layer language — contains unbounded loops). I.e. lifted code may diverge or exhaust the system's resources. In the case of the former, the reactive runtime never regains control over the evaluation of the program, never to react to new data. In the case of the latter, the program crashes. In both cases, the program becomes unresponsive, which is an undesirable state for a reactive system to be in. Furthermore, embedded RP languages often require a host program to construct the signals that make up the RP program. If this host program diverges for any reason, the program is never able to start at all, which would also not make for a responsive system.

**Reactive/Imperative Impedance Mismatch**  Reactive programs do not follow the same control flow path as traditional sequential programs. The order in which lifted functions are executed is determined by the implementation of the reactive runtime. Nonetheless, RP programs often need fine control over various side-effecting operations. E.g., operations concerning files, network communication…(i.e. IO). If these are evaluated in an incorrect order, unbeknownst to the reactive runtime, bugs may arise. Some RP languages tackle this issue by disallowing side effects in lifted functions, others by simply discouraging their use in documentation. However, this means that at the fringes of the RP program (where there is an interaction with the *outside world*), IO code is still needed to perform effects. We argue that side-effecting operations are essential and that this impedance mismatch between reactive and imperative code requires dedicated language support.

## 1.2  Contribution

To tackle the issues that plague two-layered RP languages, we propose a new language that is reactive all-the-way-through. The basic abstraction type of the language, which we call a *reactor*, serves as the sole construct for expressing (reactive) computations. Functions are absent, and therefore there is no function lifting either. E.g., the language does not have a *+ function* to compute the addition of two numbers *once* for every invocation, it only has a *+ reactor* that automatically updates *each time* one of its source signals updates. This is the first contribution of this paper. Using this language, we tackle the problems stated in Section 1.1 as follows:

- We will analyse the different groups of features in isolation to each other, with a focus on how the Reactive Thread Hijacking Problem emerges. This will allow us to get a clear grasp on how this problem emerges in RP programs, in general. If the same would be done for a two-layered RP language, not only would both layers need to be analysed separately, but also the bi-directional integration between them. These insights form the second contribution.

- By considering a pure (reactive-only) reactive language, we avoid the adverse consequences caused by the Reactive/Imperative Impedance mismatch. Without functions, reactive code and non-reactive code (i.e. imperative code) is never combined (i.e. there is no lifting), and our pure model of RP therefore lacks any form of impedance mismatch, by design. By considering the semantics of a language that is *just* a reactive language, without needing to focus on the integration with non-reactive code, we move towards a more generalised model for reactivity. I.e. something akin to the $\lambda$-calculus that captures the semantics of functional and imperative languages [41], but for reactors instead of functions. While a formal calculus that captures the essence of RP is not part of this paper, we will discuss the fundamental aspects of what makes a reactive language *reactive* in terms of our actual language. These insights into our pure model form the third contribution.

## 2 First-Order Reactors

We now present Haai, a pure reactive programming language that has been designed from first principles. Our programming model features *reactors* as the unit of (reactive) computation. The computational model of our language can be summarised as follows: reactors will, when instantiated, create the time-varying signals that make up the reactive program. And at run-time, values will be emitted by these signals to keep the program up-to-date with respect to the received input values. Functions are absent in our language model, there are only reactors. This pure model of only having reactors (i.e. Uniform Reactor Model) is familiar to other approaches in programming language design. It can be compared to the Uniform Object Model that has been popularised by languages like Smalltalk [18] and Self [50]. Like these languages, uniformity makes the design of the language simple and easy to understand.

### 2.1 Language Overview

We begin our discourse on Haai by means of a prototypical example: a unit conversion between temperature values [3]. Listing 1 presents a reactor named to-celsius. It converts incoming temperature readings in Kelvin to degrees Celsius. While the code in Listing 1 looks similar to Scheme [25] and Lisp [32] due to the use of S-expressions, it employs *reactive semantics*. By reactive semantics, we mean that reactors will need to be instantiated on time-varying signals to produce new time-varying signals. Instead of this instantiation producing a single value (e.g,. a number), it produces one or more *signals* whose value changes over time. In other words, when the reactor in Listing 1 is instantiated (e.g., on a signal containing temperature readings from a thermometer), it will keep the temperature in °C up-to-date whenever the thermometer produces a new value (in Kelvin). We will call instances of reactors *deployments*. Each

■ **Listing 1** Basic reactor definition with one source (k) and one sink.

```
1 (defr (to-celsius k)
2   (- k 273.15))
```

■ **Listing 2** Reactor definition with two sources (a and b) and two sinks (s and p).

```
1 (defr (sum-and-product a b)
2   (def s (+ a b))
3   (def p (* a b))
4   (out s p))
```

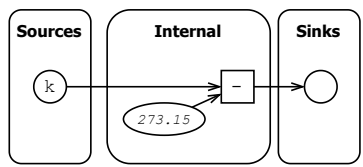
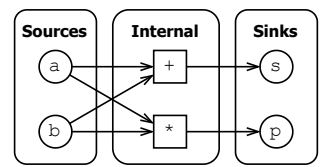

**(a)** to-celsius (Listing 1)    **(b)** sum-and-product (Listing 2)

■ **Figure 1** Reactors graphs.

deployment corresponds to the signals that were created during the instantiation of a reactor. The (- ...) expression constitutes a deployment of the - reactor.

Reactors in Haai can have more than one sink signal. An example of a reactor with two sinks is shown in Listing 2 which, given two time-varying signals (a and b), produces their sum (s) and product (p) as two separate sink signals. The out form, which is used here to denote which signals are considered as the reactor's sinks, is optional if there is only one sink, as shown earlier in Listing 1.

We visualise reactors using so-called *reactor graphs* (Figure 1). A reactor graph is a box-and-arrows diagram that — usually — consists of three regions that partition the nodes in the graph into *source*, *internal* and *sink* signals. Signals are ordered from top-to-bottom in the source and sink regions. In a reactor graph signals are drawn as ellipses or circles, and operations on them as boxes (with the name of the operator shown in the box). Note that for to-celsius (Figure 1a), the constant 273.15 is also shown as a signal: in Haai a literal is interpreted as a signal that — on a conceptual level — produces the literal value. We refer to these signals as *constant signals*.

### 2.2 Evaluation Model

Haai is a *push-based* reactive programming language [8, 12, 44]. A push-based evaluation model means that signals propagate their value directly to those that depend on them and that signals are updated whenever *any* of their dependencies change. In other words, expressions like (> x y) will update when either x changes, y changes, or when both update at the same time. This evaluation model is different to *pull-based* languages, where signals effectively pull values from their dependencies [37, 58].

⟨program⟩ ⟶ ⟨definition⟩*
⟨definition⟩ ⟶ ⟨rdefinition⟩ | ⟨sdefinition⟩
⟨rdefinition⟩ ⟶ (**defr** (⟨identifier⟩ ⟨identifier⟩+) ⟨body⟩)
⟨body⟩ ⟶ ⟨sdefinition⟩* ⟨sinks⟩
⟨sinks⟩ ⟶ (**out** ⟨expression⟩+) | ⟨expression⟩
⟨sdefinition⟩ ⟶ (**def** (⟨identifier⟩+) ⟨expression⟩) | (**def** ⟨identifier⟩ ⟨expression⟩)
⟨expression⟩ ⟶ ⟨identifier⟩ | ⟨literal⟩ | ⟨deployment expression⟩
⟨deployment expression⟩ ⟶ (⟨expression⟩ ⟨expression⟩* ...)
*Note:* The definitions of ⟨identifier⟩ and ⟨literal⟩ are considered standard.

■ **Figure 2**  Syntax rules of the Haai Language.

Like other push-based RP languages, Haai will update signals in a *glitch-free* [8] manner. A signal with multiple dependencies will only be updated if all the dependencies are current. For example, the expression (> (+ x 1) x) produces a signal whose value should always equal to true. When x changes, the update of > will be postponed until the update of the + signal has also been propagated. We will refer to each instant of the logical clock that drives the propagation of values in the program as an *update turn*, or *turn* for short. This terminology has been inspired by actor systems, which often process messages in a turn-based manner [13].

In brief, one can view the evaluation model of a Haai program as a perpetually-running loop that constantly pushes data (e.g., from external sources) into the sources of the reactive program. The values pushed to these sources will cause a cascade of signal updates that eventually reaches the sinks of the program.

### 2.3 Syntax

The syntax rules of Haai are shown in Figure 2. In brief, a Haai program consists of reactor definitions (which use the **defr** keyword) and signal definitions (which use the **def** keyword). The latter form supports multiple variable declarations (e.g., as in Listing 2) to support deployments of multiple-sink reactors.[1] Reactor definitions contain (internal) signal definitions, and a definition of sinks. When a reactor has only one sink signal, the **out** form can be omitted, the last expression in the body will then determine the single sink signal. Expressions are either literals (which correspond to constant signals), variables (which refer to signals or reactors by name) or deployment expressions (which combine other expressions). Deployment expressions are similar to application expressions in non-reactive languages: they combine an operator with operands. Note that no syntax is provided for conditionals, yet. Conditional signals will be discussed in Section 4.1.

---

[1] The **def** form in Haai is thus similar to **define-values** in Racket [17].

■ **Listing 3**  A simple reactive program which checks whether the temperature readings of a network-connected thermometer are below the freezing point of water.

```
1 (def temperature (ws-in "localhost:3333"))
2 (def freezing-temperature (negative? (to-celsius temperature)))
3 (def output (ws-out "localhost:4444" freezing-temperature))
```

### 2.4 Standard Library of Reactors

In the absence of lifting, the Haai language has a set of built-in primitive reactors that enable programs to perform basic computations on numbers, strings, and basic data structures (pairs and vectors), similar to Scheme [25]. To avoid the Reactive/Imperative Impedance Mismatch only primitive procedures that are referentially transparent have a reactor counterpart. There are no reactors that can perform side effects (e.g., vector-set!), perform IO (e.g., read), evaluate data (e.g., eval) or are involved with continuations (e.g., call/cc). The obvious difference between these procedures in Scheme and their reactor counterparts is that reactors will continuously re-run their computation whenever a signal changes. Besides these reactors, Haai features dedicated built-in reactors that are more suitable in a reactive setting, we discuss these at relevant points in the paper.

### 2.5 Producer and Consumer Signals

To allow Haai programs to react on values from the outside world, we propose a set of specialised reactors that produce signals that can act as (external) data producers. Vice versa, we also propose a set of specialised reactors that processes — external to the RP program itself — the output values produced by the reactive program. These two groups of reactors make it possible for Haai programs to perform IO, without those programs needing to implement IO themselves.

We explain these reactors by means of an example. Listing 3 contains a basic reactive program that determines whether a thermometer is reading freezing temperatures. On Line 1 the ws-in reactor is deployed, which is a *Data Producing Reactor*. Its sole argument specifies an address of a WebSocket server in which the reactive program will be able to connect to such that it receives the data that it broadcasts. We thus make it the responsibility of the language runtime (i.e. the interpreter) to actually maintain that WebSocket connection, without requiring a programmer to write any imperative code in their reactive program. This is similar to primitive methods in Smalltalk [18] which are internally annotated to be evaluated using native (VM) code, transparent to the Smalltalk user. On Line 3 the the ws-out (a *Data Consuming Reactor*) is used which does the opposite. Its sole argument specifies the address of another WebSocket server in which the reactive program can send data to. The exact application protocol used by these primitives is irrelevant to the discussion of this paper, and is therefore not discussed.

These special reactors provide a clear separation between sensing and actuating (i.e. with side effects in the real world) and the reactive program (i.e. the program with reactors and signals). The main idea here is that future language implementations can provide domain-specific operators: e.g., a variation of Haai for embedded devices can provide specialised operators for the various sensors and actuators in a physical device (similar to from_topic and to_topic in RxROS [27] for ROS [29]). In the rest of this paper, we will not consider these specialised operators, as they are not essential to Haai's semantic model.

## 2.6 Reactivity Guarantees

In earlier work [53], we defined three different levels of reactivity guarantees that classify a RP program (or language's) ability to process all incoming events. These are related to the *Reactive Thread Hijacking Problem* from Section 1.1. We briefly summarise these three levels below:

**Strongly reactive**  A strongly reactive programming language enforces a constant upper bound on the evaluation time of each turn. In other words, regardless of the values that flow over the reactive program's signals: a constant upper-bound can be determined statically on the evaluation time of each turn. Strong reactivity is a desirable property for (soft) real-time systems.

**Eventual reactive**  An eventual reactive programming language is one for which a constant upper bound does not exist, but one that can guarantee that every turn will terminate in finite time. As a consequence, an eventual reactive program might appear temporarily unresponsive.

**Weakly reactive**  A weakly reactive programming language is one which cannot provide any guarantees at all on the evaluation time of each turn. A turn in a weakly reactive language might diverge and never terminate. I.e., the reactive evaluation thread might get hijacked, resulting in a program that can forever be unresponsive.

A reactive programming language is a language that is used to make reactive applications. I.e. applications that respond in a timely manner to any event that triggers a re-computation by letting each turn run to completion. Therefore, we claim that a reactive programming language should be carefully designed such that it can guarantee the right level of reactivity needed for the targeted application. The basis of Haai is — as we will discuss in Section 2.7 — strongly reactive. In later sections of this paper, we will present various extensions, and we will analyse for each (group) of) feature(s) the impact on this classification. An overview of features with their classification is shown near the end of this paper (Table 1).

## 2.7 Reactivity Guarantees of First-Order Haai

We now discuss first-order Haai's reactivity classification. We start by considering what a Haai program looks like, and how it behaves at run-time. A first-order Haai program consists of a finite set of signal and reactor definitions. Each signal definition deploys the reactor(s) that it contains, which result in a (cascade of) deployments of the constituent reactors. Without recursive reactor definitions,[2] the hierarchy of deployed reactors follows the syntactical structure in the source code. I.e., while reactors can be composed of other reactors, each reactor has access to a smaller number of possible constituent reactors. Given this structure of reactor deployments, a first-order Haai program will, once deployed, always consist of the same (number of) deployments. And as each deployment has a fixed set of signals, the total number of signals also remains constant. Therefore, propagating values in each turn will — in a worst-case scenario where all signals are affected by a change — take a constant amount of time, which is a requirement for a strongly reactive language.

Unfortunately, some built-in reactors have a computational complexity that is not constant (i.e. not in $\mathcal{O}(1)$). For example, reactors that like their Scheme counterparts operate on data structures of arbitrary sizes or contain (as part of their implementation) a form of iteration. If these reactors are considered as being part of the standard library, we consider first-order Haai to be an *eventually reactive language*, instead of strongly reactive one. We have therefore limited the standard library in Haai to reactors that can be implemented to have a $\mathcal{O}(1)$ time complexity, and claim that this makes the language strongly reactive. This restriction does limit expresiveness, a design decision that will be further discussed in Section 8.

## 3 Stateful Reactors

State is a fundamental aspect of an RP language. Not only do RP languages typically retain the most recent value produced by a signal to incrementally evaluate turns, RP languages often provide a mechanism to delay a signal: i.e. to get access to the past value(s) in later turns. For example, an application that processes real-time sensor values may have to apply damping to reduce jitter, or a program may only need to produce an alert if a certain condition has been sustained for a certain duration. In these applications, the behaviour of the program can no longer be expressed from just the *current* values.

### 3.1 Trampoline Variables

Typically, an RP language provides stateful operators that hide a stateful variable in their implementation. Examples include integral [9, 15], pre [58], latch [54], and foldp [8, 12, 34]. We propose a different approach that makes mutable state an explicit part of the language model. In Haai, deployments are able to accumulate state by means of so-called *trampoline variables* — or trampolines for short. They provide programmers with a general notion of state in the RP model. In essence, a trampoline is a variable, local to each deployment, that is updated after each turn.

---

[2] Recursive reactors will be discussed in Section 5.

■ **Listing 4**  Reactor that retains the minimum and maximum value of a signal using trampolines.

```
1  (defr (min-max s | (i s) (a s))
2    (def mi (smallest s i))
3    (def ma (largest s a))
4    (out mi ma | mi ma))
```

...

⟨rdefinition⟩ ⟶ (defr (⟨identifier⟩ ⟨identifier⟩⁺ [ | (⟨identifier⟩ ⟨expression⟩)⁺ ] ) ⟨body⟩)
⟨sinks⟩ ⟶ (out ⟨expression⟩⁺ [ | ⟨expression⟩⁺ ]) | ⟨expression⟩

■ **Figure 3**  Changes made to the syntax rules from Figure 2 to support stateful reactors.

The semantics of trampoline variables is best understood with an example. Listing 4 shows a reactor min-max that keeps track of the minimum and maximum values observed of a numerical signal. The state maintained by this reactor is the minimum and maximum values observed so far: i.e. its trampoline variables. They are defined next to the source signals, a vertical bar (|) separates the formal parameters from the trampoline variables. Each trampoline variable has a name and an expression that determines its initial value. For min-max, both trampoline variables are defined by taking the *current*, at-deployment time, value of the source signal. From the value stored in the trampoline variable and the current value of the source signal, the new minimum and maximum can be computed: i.e. the signals mi and ma. These two signals are used both as sink signal of the min-max reactor, and as the signals that update the trampolines at the end of the turn. Another vertical bar separates the ordinary sinks (i.e. those that will be made available to the deployment-site) from the expressions that denote the new value of the trampoline variables (they denote an internal assignment managed by the language).

Figure 3 shows the modified syntax rules for defining reactors with trampolines, as used in Listing 4. Square parenthesis denote optional parts here.

Trampoline variables are always updated at the end of the turn, hence providing access to an old value of a signal in a later turn. This update does not cause an immediate reaction (i.e. a new turn). It is only when a reactor deployment is updated in reaction to a change to a source signal that the value of the trampoline variable is propagated.

Figure 4 visualises min-max. Trampolines are contained in a new region and have two incoming arrows. The left arrow identifies the signal that initialises the trampoline. The bottom arrow identifies the signal that updates the trampoline. Note that trampolines may introduce cycles in the reactor graph. These cycles should disappear when the the (delayed) update arrow is omitted.

Trampolines can be used to implement trampoline reactors that have similar behaviour to the stateful operators of other RP languages. Listing 5 implements pre, an operator that delays the value of a signal by one turn. Each time a source signal s

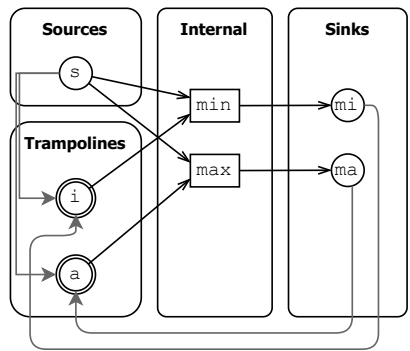

**■ Figure 4**  Reactor graph of min-max (from Listing 4).

**■ Listing 5**  Trampoline reactor that delays a signal by one turn.

```
1  (defr (pre s init | (acc init))
2    (out acc | s))
```

produces a value, (pre s init) produces the previously-produced value of s (except the very first time when the value of the init signal is produced). Other stateful operators, mentioned earlier in this section, can be implemented similarly.

### 3.2 Impact on Reactivity Guarantees

Trampoline variables do not impact the reactivity guarantees. While a program may consist of many trampolines, the total number of trampolines is fixed (similar to how the total number of signals is fixed). Therefore, at the end of each turn, the same amount of work needs to be performed to keep these trampolines updated (i.e. taking the value of the indicated update signals and storing it for the next turn). This makes stateful reactors in Haai a *strongly reactive* feature.

**Compared to two-layered RP languages**  Often, RP languages contain one or more operators that requires a function to update a stateful signal: e.g., Elm's [12] foldp. The function supplied to foldp is applied (using base-layer semantics) each time the incoming signal is updated. If these functions are unchecked, as is the case in Elm, the RP language becomes weakly reactive as the function supplied to foldp may (accidentally) diverge.

## 4    Higher-Order Reactors

The introduction of higher-order reactors means that some deployments become polymorphic. As such the dependencies of signals may change at run-time. Before we discuss these polymorphic deployments, we first discuss *conditional signals* as they provide a gentle introduction to signals with dynamic dependencies.

...
⟨expression⟩ → ... | ⟨conditional expression⟩
⟨conditional expression⟩ → (if ⟨expression⟩ ⟨expression⟩ ⟨expression⟩).

■ **Figure 5** Changes made to the syntax rules from Figure 2 to support conditional signals.

■ **Listing 6** A Haai program that computes the next number in a Collatz sequence.

```
1  (defr (collatz-step n)
2    (if (even? n)
3        (/ n 2)
4        (+ (* n 3) 1)))
```

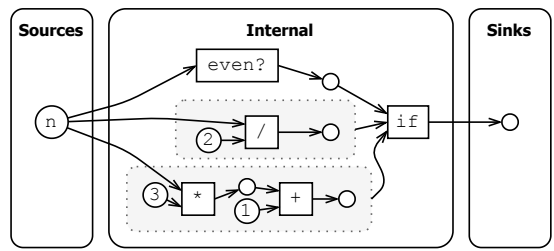

■ **Figure 6** Reactor graph of collatz-step (from Listing 6).

## 4.1 Conditional Signals

In Haai, conditional signals are signals whose value is determined in relation to a specified condition. Depending on the value of the (time-varying) condition either the value of one signal or another signal will be produced. At run-time, conditional signals will *toggle* their active dependencies between two possible states, depending on the value produced by the conditional signal. Conditional signals are created using the if form whose syntax is standard, but shown nonetheless in Figure 5.

Listing 6 shows a reactor that uses the if form to reactively compute the next number in a Collatz sequence [26]. Depending on the parity of n, (collatz-step n) produces either the values of (/ n 2) or (+ (* n 3) 1). The reactor graph of collatz-step is shown in Figure 6. The signal nodes that correspond with the consequent and alternate expressions of the if are grouped together by a grey box, to highlight that all signals contained therein either react together, or not at all.

An important aspect of if is that either the signals in the consequent or alternative expression are instantiated during the deployment of a reactor that uses if. It is only when the truth value of the conditional signal changes, in a later turn, that the other signal(s) will be instantiated. This makes the semantics of if equivalent to a more general approach to programs with dynamic signal dependencies, which we now discuss in more detail.

■ **Listing 7**   A Haai program that uses higher-order deployment expressions.

```
1  (defr (temp-locale time temp)
2    (def r (if (even? time)
3                to-celsius
4                to-fahrenheit))
5    (r temp))
```

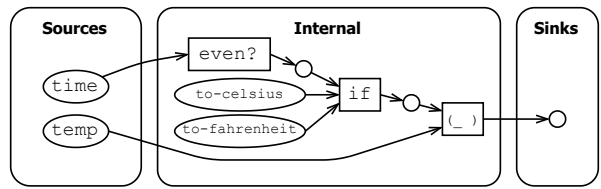

■ **Figure 7**   Reactor graph of `temp-locale` (from Listing 7).

## 4.2  Dynamic Deployments

We now make reactors in Haai first-class values: i.e. they can now be emitted by signals. We will call signals that only emit reactors *reactor signals*. The semantics of deployment expressions changes in a remarkable manner, if a reactor signal is used in operator position: deployment expressions become polymorphic. This gives rise to a phenomenon called *dynamic deployments*: deployments created while the program is running. An example of a reactor (`temp-locale`) in which a dynamic deployment occurs is shown in Listing 7. On Line 2, a reactor signal r is defined whose value is either equal to the `to-celsius` reactor or the `to-fahrenheit` reactor. This signal is then used in operator position on Line 5 which causes the program to toggle between deployments of these two reactors. As a consequence, the signal that the `temp-locale` reactor created, continuously toggles between the sink signal of these two deployments. Note here that signals are not first-class values.

A reactor graph of `temp-locale` is shown in Figure 7. Note here that a consequent or alternate expression consisting of single named signal or reactor is not contained in a grey-coloured box. These signals (reactors) already exist and are updated independent of the conditional signal.

Semantically, a reactor deployment is only active when needed by a dynamic deployment. I.e. during the very first turn either the `to-celsius` or `to-fahrenheit` reactor will be deployed, depending on the parity of `seconds`. One second later, when the parity of `seconds` changes, the other reactor will be deployed and the other deployment will be *deactivated*. Another second later, the previously-allocated deployment will be re-activated and the other one will be deactivated. This type of switching is not a new idea [11], but differs from most other RP languages. E.g., Yampa [37] and FrTime [8] will disconnect signals if they are replaced by other signals. In these languages, signals disconnected by switching are removed from memory by relying on a garbage collector (as provided by their respective host languages).

### 4.3 Impact on Reactivity Guarantees

Conditional signals and dynamic deployments result in a language that is still *strongly reactive*. Both features allow for the creation of reactive programs where the amount of signals may change depending on the number of deployments that have been activated. These deployments may be created dynamically and thus the amount of work a single turn may take is not constant, as it depends on actual deployments that are active of the dynamic deployments. Nonetheless, an upper-bound can still be calculated for each possible path by analysing which reactor values flow over the reactor signals (e.g., by means of a flow analysis).

**Compared to two-layered RP languages**    In general, we have identified two categories of RP languages that support signals with dynamic dependencies. Some languages (e.g., Elm [12]) do not allow for the run-time creation of new signals: a conditional signal can then only toggle between two existing signals. We would consider conditional signals in those languages also as strongly reactive. Other RP languages allow for new signals to be created while the program is running: either by allowing a signal graph to be created (e.g., Yampa [37] and Dunai [40]) and be swapped with some operator, or by allowing (lifted) code to create new signals incrementally (e.g., REScala [44]). We consider these languages weakly reactive as this creation is driven by unchecked (base-layer) host code.

## 5 Recursive Reactors

Existing RP languages often have support for two different kinds of recursion. The first kind, which we call *lifted recursion*, occurs when a recursive function is lifted. A lifted recursive function can be applied on a signal and the reactive runtime uses the semantics of the base layer to evaluate the lifted function. The second kind, which we call *graph recursion*, occurs when the recursive (or looping) abilities of the host language are used to recursively generate signals.[3]

### 5.1 Graph Recursion in Haai

In the absence of functions, the only supported form of recursion in Haai is graph recursion. Surprisingly, (graph) recursion in Haai is simple to understand as it behaves the same as regular recursion in other programming languages: i.e. by using self-reference. Instead of functions that can apply themselves at reaction-time, reactors in Haai can deploy themselves, providing a base case to stop recursive deployments (e.g., using if). An example of a recursive reactor is shown in Listing 8. It defines a

---

[3] Some languages also consider stateful operators as a kind of recursion [37, 58]. However, unlike lifted and graph recursion, this form of recursion does not affect the evaluation of a single turn. We therefore do not consider it as a form of recursion in this section.

■ **Listing 8** Reactor that constructs a recursively-defined chain of signals, depending on a given number.

```
1  (defr (collatz-length num count)
2    (if (= num 1)
3      count
4      (collatz-length (collatz-step num) (+ count 1))))
```

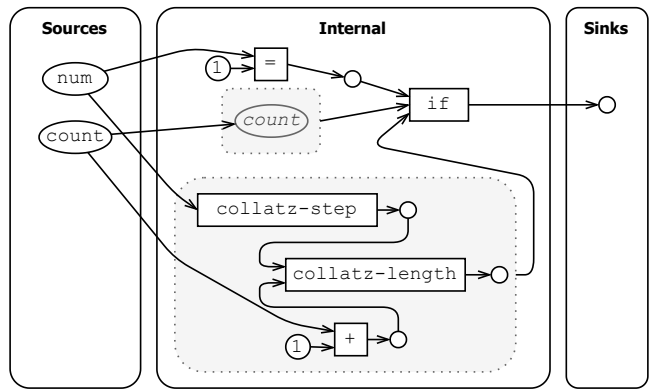

■ **Figure 8** Reactor graph of collatz-length (from Listing 8).

recursive reactor named collatz-length to reactively compute the length of a Collatz sequence. Figure 8 shows the reactor graph of collatz-length. The recursive deployment of collatz-length is included in the second grey-coloured rectangle. It is only active, and thus deployed, in a turn when num is not equal to 1.

We admit that the use of graph recursion is quite atypical for this example. In a two-layered reactive programming language, the same program would usually be implemented by lifting a (recursive) function. A more practical example of graph recursion is constructing reactive sorting networks [38].

### 5.2 Impact on Reactivity Guarantees

Recursion is — as in any programming language — not without any risks: deployments of recursive reactors may also diverge. This makes recursion a *weakly reactive* language feature.

There are some subtleties regarding reactive recursion, specific to RP. We discuss these now. First, note that a program like (defr (loop t) (loop (+ t 1))) will never be able to be fully deployed. Deploying (loop time) causes the creation of an infinite chain of signals that all transitively depend on time (each signal incrementing the value of the previous signal in the chain by one). The same cannot be said of (defr (loop2 t) (if (= t 100) 0 (loop (+ t 1)))). If (loop2 time) is deployed, it will — like before — construct chain of deployments, except that this time the size of the chain will be limited in the presence of a recursive base case. However, if the time signal's value itself becomes bigger than 100 in a later turn, the program will create an infinite deployment loop

...
⟨expression⟩ → ... | ⟨rho expression⟩
⟨rho expression⟩ → (rho (⟨identifier⟩$^+$...) ⟨body⟩).

■ **Figure 9** Changes made to the syntax rules from Figure 2 to support anonymous reactors.

as the base case will then never be reached. In other words, depending on the values produced by the time-varying signals that populate an RP program, a recursive deployment may diverge. This may happen in any turn, not just the first.

**Restricting Recursion** To avoid programs to become weakly reactive, the use of recursion needs to be restricted (at the cost of expressiveness). One approach could be to apply techniques from termination detection systems [6, 28] such as primitive and structural recursion [56], or size change termination [6, 36]. Note that termination in this context does not mean termination of the RP program in its entirety — which would be undesirable — but the termination of the deployment loops that recursive reactors may manifest. This should guarantee a system that is at least eventually reactive. An overview of the integration of termination checkers, to ensure eventual reactivity in Haai-with-recursion, is outside the scope of this paper.

**Compared to two-layered RP languages** We have already — at many points in this paper — discussed how two-layered languages can make an RP program weakly reactive, and will not do so again. We do, however, note the following: while at first sight only supporting graph recursion in Haai looks like a fundamental limitation, programs that would in other RP languages make use of lifted recursion can still be expressed (which will be discussed in more detail in Section 8.3). One fundamental difference to note here though — with respect to two-layered RP languages — is that it is pure reactive code in our model that is responsible for the recursive creation of signals. In two-layered RP languages, this is often the responsibility of (lifted) base-layer code [8, 44].

## 6 Anonymous Reactors

Similar to functional languages, Haai has a construct to make *anonymous* abstractions for computations.

### 6.1 Reactors with Scope

Unlike functional languages, which usually use the $\lambda$ symbol or `lambda` keyword, we denote anonymous reactors using $\rho$ or `rho`, to emphasise that they create reactors, not functions. Figure 9 presents the modified syntax rules.

■ **Listing 9** A reactor definition with an anonymous reactor definition embedded within.

```
1  (defr (make-temp-locale in-celsius)
2    (rho (temp)
3      (if in-celsius
4        (to-celsius temp)
5        (to-fahrenheit temp))))
```

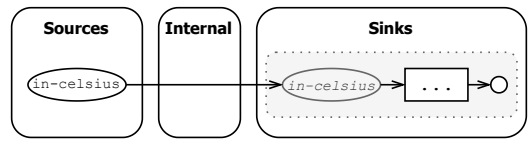

■ **Figure 10** Simplified reactor graph of make-temp-locale (from Listing 9): the internal structure of the capture has been omitted.

Anonymous reactors in Haai are lexically scoped. As a consequence, deployments of anonymous reactors have two types of sources: explicit sources (i.e. sources that are passed at deployment-time) and implicit sources (i.e. sources captured when the encompassing reactor was being deployed). The use of anonymous reactors is exemplified by Listing 9. The implementation of make-temp-locale shows how a reactor creates an anonymous reactor. The in-celsius source signal determines whether to convert temperatures to either °C or °F. The make-temp-locale reactor itself emits (on a constant signal) a representation of this anonymous reactor (a *capture*, c.f. a closure) that has this signal in scope. Deployments of that capture will react accordingly whenever in-celsius changes (even if it is not part of the reactor's explicit sources).

The reactor graph of make-temp-locale is shown in Figure 10. The grey-coloured box represents the rho expression. We used this visual notation earlier in Figures 6 and 8 for if. Like if, rho delays the deployment of certain signals. Unlike if, however, is that the signals contained therein are not necessarily deployed by the same deployment.

Remember that signals are not first-class. To pass around signals, an expression like (rho () x) can make a (first-class) run-time representation of an x signal. To use this signal elsewhere, this capture needs to be deployed. In other words, captures make it possible to store signals in data structures, not just the values that they produce. This approach results in a general model of higher-order reactive programming, without any specialised higher-order operators.

## 6.2 Impact on Reactivity Guarantees

Deployed captures are like ordinary reactors, except that they have additional signals in scope (i.e. captured signals). As such, they should not directly affect the language's reactivity tier. We could therefore prematurely claim that anonymous reactors are a *strongly reactive* language feature, but unfortunately that is not the case. Similar to non-reactive languages, a recursive fixpoint combinator can be implemented. Listing 10 provides a possible definition of a recursive fixpoint reactor. It closely resembles

■ **Listing 10**   fix: a recursive fixpoint reactor.

```
1  (defr (fix f)
2    ((rho (x) (f (rho (y) ((x x) y))))
3     (rho (x) (f (rho (y) ((x x) y))))))
```

the call-by-value Y-combinator [42]: the only difference being the use of `rho` instead of `lambda`. This reactive fix is similar to a fix in a call-by-value language: the expansion occurs for every new recursive deployment, allowing for self-deployment (cf. self-application), which has similar consequences as ordinary recursion, by self-reference as discussed in Section 5.2. This makes anonymous reactors a *weakly reactive* language feature, quite a degradation with respect to our earlier assumption. When the creation of Y-combinators is restricted (e.g., by forgoing the capture semantics, or by an occurs check to find recursive type definitions on reactors, i.e. the simply typed lambda calculus) anonymous reactors should not affect the reactivity classification. Under these conditions, we can classify `rho` as a *strongly reactive* feature.

**Compared to two-layered RP languages**   Anonymous abstractions in RP are not a completely new idea. Signal functions, as introduced by Yampa [37], are closely related to (anonymous) reactors. However, unlike anonymous reactors, Yampa's signal functions cannot capture signals. Furthermore, signal functions in Yampa can be created by the evaluation of non-reactive (base-layer) code.

## 7   Implementation

The Haai language has a prototypical implementation developed in Racket [16]. An interactive interpreter loads and runs code provided by a programmer, akin to a Read-Eval-Print Loop. Reactor definitions are loaded as entries in a so-called reactor table. Signal definitions are immediately deployed (i.e. *global deployments*) and automatically react to changes. The implementation employs an asynchronous — thread-based — evaluation model. One thread is responsible to sense the environment for new data of any real-world sources, another thread is responsible to read expressions entered by the programmer and a third thread is responsible for performing update turns (i.e. to propagate changes received by the other two threads: i.e. to either instantiate new signals or propagate values through signals created earlier).

The prototypical implementation has support for various kinds of Data Producing and Consuming Reactors (Section 2.5). Besides WebSockets and other network-based communication protocols the language provides simple primitives to build GUI widgets, inspired by [12, 24]. In the future, we aim to extend our prototypical interpreter to support more kinds of application environments.

■ **Table 1** Overview of the features of Haai, and which effect they have on the classification of the language according to its reactivity guarantees.

| Type of Reactors | Reactivity Level |
|---|---|
| First-Order Reactors (Section 2) | Eventual or Strong∗ |
| Stateful Reactors (Section 3) | Strong |
| Higher-Order Reactors (Section 4) | Strong |
| Recursive Reactors (Section 5) | Weak |
| Anonymous Reactors (Section 6) | Weak or Strong† |

∗ Strong if the primitive reactors are restricted to those that can update their sink(s) in $\mathscr{O}(1)$ time.

† Strong if programs with recursive fixpoint combinators are statically rejected. The act of capturing a signal and depending on it elsewhere — the main intent — is not a weak feature.

## 8 Evaluation

In this section, we discuss how Haai tackled the issues identified in two-layered RP languages, and compare its level of expressivity to those languages.

### 8.1 Reactive Thread Hijacking Problem

An overview of every language feature and their corresponding reactivity level classification is shown in Table 1. It shows that most of Haai's features are — assuming a first-order base language where all primitive reactors update sinks in $\mathscr{O}(1)$ complexity — strongly reactive. In other words, programs written using (only) these features result in a program that is guaranteed to be responsive: each turn has a fixed (constant) upper-bound on the number of operations needed to update all signals, regardless of the actual values propagated by these signals. The cost of an RP language being strongly reactive is that some programs might no longer be expressible. Nonetheless, we have identified various other strongly reactive languages (even ones that are not strictly speaking RP languages) which shows that — at least for some problem domains — strong reactivity surmounts expressivity. We discuss such languages in Section 9.

### 8.2 Reactive/Imperative Impedance Mismatch

The specialised Data Producing and Data Consuming Reactors — as the only means to perform effects in a pure language — mitigate the adverse consequences that may emerge from the Impedance Mismatch. These operators drive the interaction between the world of pure reactors and the real world of sensors (which can serve as sources of the RP program) and actuators (which take the values produced by the sinks of the RP program). These operators are — in practice — only used in the top-level, resulting in a clean separation between the reactive world and the outside world.

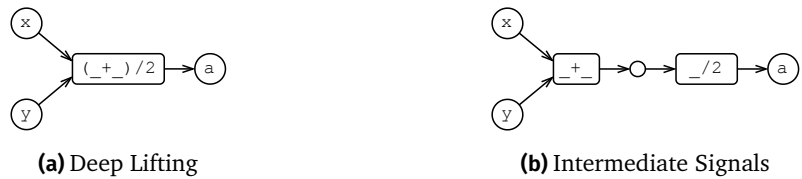

**(a)** Deep Lifting            **(b)** Intermediate Signals

■ **Figure 11** Signal graphs that compute the average of two signals. Both a signals are semantically equivalent. Constants have been inlined in these figures for simplicity.

## 8.3 Expressivity of a Pure Reactive Language

The Haai Language does not feature functions, only reactors. This raises the question whether or not this limits expressivity. If we ignore RP programs with embedded imperative side effects, we claim that there is no fundamental restriction with regards to expressivity. We base this claim on an observation found in two-layered reactive programming languages. In these languages, there is no observable difference between a signal created by a lifting a composed function, and a signal that depends on intermediate signals that correspond with the internal structure of the original function. The designers of FrTime discussed this too in the context of *deep lifting* [7], which is an optimisation technique that "eliminate[s] the large dataflow graphs that lifting would otherwise create". Figure 11 visualises the ideas behind deep lifting. Figure 11a shows a signal that directly computes the average of two other signals and Figure 11b uses an intermediary signal that first computes their sum. Both resulting signals are semantically equivalent.

In the case of Haai, only programs with intermediary signals (as in Figure 11b) can be implemented. To show that expressiveness is not limited, we have to show that programs that use composed functions can also be expressed with intermediary signals. For simple programs involving only lifted functions, the composed function can be decomposed into it constituents. However, functions may be used in different ways besides plain lifting. Consider FrTime's collect-e operator [8]. collect-e takes a procedure as an argument that is then applied to the values produced by a signal and the current value of an accumulator. Operators like collect-e apply said procedures *not* on the signals themselves, but on the values produced by signals. In most cases, there is no difference between a function being applied to the values of a signal, or to the signal's value being passed to the procedure. Only primitive functions require the actual value of a signal [7]. Such operators are absent from Haai, but are, effectively, a trick needed to sample a signal's value for a stateful accumulation (which are the semantics of trampoline variables). As such, we conclude that expressivity in Haai is comparable to two-layered RP languages.

## 9 Related Work

In this section, we present an overview of languages (RP or otherwise) that tackle similar problems.

### 9.1 Reactive Thread Hijacking Problem

We now discuss languages that we would classify as being either eventually reactive or strongly reactive. For brevity, weakly reactive languages are not discussed.

**Reactive Programming** RT-FRP [58] is a functional reactive programming (FRP) language that we consider to be *strongly reactive*. The language has support for lifting and multi-modal signals (i.e. switching). To ensure strong reactivity, the language only allows functions from the simply-typed lambda calculus (i.e., without recursion, nor self-application [41]) to be lifted. The semantics of E-FRP [57] are expressed in terms of RT-FRP. We therefore also consider E-FRP to be *strongly reactive*.

EmFRP [46] is a reactive programming language that targets embedded devices. We also consider it to be *strongly reactive*. The language has no support for recursion (neither lifted recursion nor graph recursion). A reactive program consists of a static set of signal definitions and supports no switching.

The Stella language [51, 53] has a reactive language embedded within an actor-based one. At run-time, the Stella implementation checks that these lifted functions terminate (by performing dynamic size change termination [36]). The reactive language of Stella does not support switching: the internal dependencies between signals can be derived statically. However, there is support for replication: parts of a program can be replicated, where each copy is connected to a different element in a (time-varying) set. As these sets are constrained to finite sets, this replication is bounded [52]. We therefore consider Stella to be an *eventually reactive* language.

ActiveSheets [22, 54] is an RP language contained within a spreadsheet program. ActiveSheets supports cells connected to time-varying data streams, and changes to these cells are propagated automatically. Most primitive operations in ActiveSheets have a worst-case time complexity of $\mathcal{O}(1)$. However, some spreadsheet operations (e.g., those that *search* like VLOOKUP) are not. Nonetheless, these operations are guaranteed to terminate, making ActiveSheets *eventually reactive*.

There exist also various formal systems that aim to prove liveness (i.e. eventual reactivity) for reactive programs and streams [2, 47]. The main difference is that these models assume a streaming model, where programmers are explicitly in control over how a stream evolves over time (by means of a functional program), similar to streams in [1]. This model is different from RP presented in this paper, where signals compute their value in terms of the values of their dependencies.

**Synchronous Programming** The ideas of (strong) reactivity can also be found in other paradigms: e.g., Synchronous Programming (SP) [19]. We discuss two SP languages that closely resemble Haai: Lustre [20] and Lucid Synchrone [5]. Similar to our language, programs expressed in these languages consist of dataflow graphs that express how each time-varying value is computed. Signals in both Lustre and Lucid Synchrone have statically encoded timing information ("a clock") which is used to determine which signals can be composed together (i.e. applying a simple function like + on two signals that do not share the same clock is disallowed). This is the main difference between SP and RP. In RP, signals are always updated with the current values of their dependencies.

### 9.2 Reactive/Imperative Impedance Mismatch

To solve the issues of the Reactive/Imperative Impedance mismatch, reactive code must clearly be separated from imperative code. As Haai is — to the best of our knowledge — the first pure reactive language, we list languages here where this separation has been made explicit by other means.

In Yampa [37], signal functions are created by the application of *pure* functions that are unable to perform side effects. To run a Yampa program, a signal function needs to be combined with IO actions that provide inputs (sense) and perform actions with the outputs (actuate), using a function named the reactimate [10]. I.e. reactimate returns an IO action that is the actual RP program. Therefore, a working Yampa application contains both reactive and imperative (IO) code. FRP anguages inspired by Yampa, such as SFRP [4] and Dunai [40], share this style of combining reactive and IO code. Besides these languages, we also note two languages that target embedded applications that, like Yampa, have an explicit separation between RP and IO code. EmFRP [46] compiles reactive programs into C/C++ code which needs to be supplemented with the actual driver (IO) application before the application (as a whole) can be subsequently compiled. Juniper [21] allows C/C++ code to be inlined in the Juniper program itself, such that it can be compiled into a full program.

Finally, the aforementioned Stella language [53] proposes a clear separation between reactive and imperative code by only allowing actors to perform side effects, and reactors to be pure. A working Stella program also requires both imperative and reactive code, which means that Stella is not a pure language either.

## 10    Conclusion

This paper presented Haai, a reactive programming language designed from first principles. Inspired by Self and Smalltalk's Uniform Object Model, Haai employs a Uniform Reactor Model where programs are composed of reactors and every runtime value is passed over signals spawned by deploying these reactors. We presented Haai in a layered approach. Section 2 presented Haai's first-order language model. Reactors are the computational abstraction of Haai to denote dependencies between signals. Sections 3 to 6 then presented various extensions. For each extension we discussed the effects they had on so-called reactivity guarantees, i.e. how strict the language is to enforce a certain degree of responsiveness of programs written in the language with the discussed features. Certain powerful features were identified in RP languages that hamper a language's ability to enforce the right degree of responsiveness. While these issues are also present in our language, our pure model allowed us to abstract over the integration of the reactive code in a non-reactive model, and consider each feature in isolation. This allowed us to increase our understanding of the RP paradigm in general.

**Acknowledgements**  We would like to thank the anonymous reviewers for their constructive comments. Bjarno Oeyen is funded by the Research Foundation - Flanders (FWO) under grant number 1S93822N.

## References

[1] Harold Abelson and Gerald J. Sussman. *Structure and Interpretation of Computer Programs, Second Edition*. MIT Press, 1996. ISBN: 0-262-01153-0.

[2] Patrick Bahr, Christian Uldal Graulund, and Rasmus Ejlers Møgelberg. "Diamonds are not forever: liveness in reactive programming with guarded recursion". In: *Proceedings of the ACM on Programming Languages* 5.POPL (2021), pages 1–28. DOI: 10.1145/3434283.

[3] Engineer Bainomugisha, Andoni Lombide Carreton, Tom Van Cutsem, Stijn Mostinckx, and Wolfgang De Meuter. "A survey on reactive programming". In: *ACM Computing Surveys* 45.4 (2013), 52:1–52:34. DOI: 10.1145/2501654.2501666.

[4] Guerric Chupin and Henrik Nilsson. "Functional Reactive Programming, restated". In: *Proceedings of the 21st International Symposium on Principles and Practice of Programming Languages, PPDP 2019, Porto, Portugal, October 7-9, 2019*. Edited by Ekaterina Komendantskaya. ACM, 2019, 7:1–7:14. DOI: 10.1145/3354166.3354172.

[5] Jean-Louis Colaço, Grégoire Hamon, and Marc Pouzet. "Mixing signals and modes in synchronous data-flow systems". In: *Proceedings of the 6th ACM & IEEE International conference on Embedded software, EMSOFT 2006, October 22-25, 2006, Seoul, Korea*. Edited by Sang Lyul Min and Wang Yi. ACM, 2006, pages 73–82. DOI: 10.1145/1176887.1176899.

[6] Byron Cook, Andreas Podelski, and Andrey Rybalchenko. "Terminator: Beyond Safety". In: *Computer Aided Verification, 18th International Conference, CAV 2006, Seattle, WA, USA, August 17-20, 2006, Proceedings*. Edited by Thomas Ball and Robert B. Jones. Volume 4144. Lecture Notes in Computer Science. Springer, 2006, pages 415–418. DOI: 10.1007/11817963_37.

[7] Gregory Harold Cooper. "Integrating Dataflow Evaluation into a Practical Higher-Order Call-by-Value Language". https://web.archive.org/web/20230509111949/https://cs.brown.edu/people/ghcooper/thesis.pdf, last accessed on 2023-05-09. PhD thesis. Providence, Rhode Island, USA: Brown University, May 2008. ISBN: 978-0-549-89696-8.

[8] Gregory Harold Cooper and Shriram Krishnamurthi. "Embedding Dynamic Dataflow in a Call-by-Value Language". In: *Programming Languages and Systems, 15th European Symposium on Programming, ESOP 2006, Held as Part of the Joint European Conferences on Theory and Practice of Software, ETAPS 2006, Vienna, Austria, March 27-28, 2006, Proceedings*. Edited by Peter Sestoft. Volume 3924. Lecture Notes in Computer Science. Berlin, Heidelberg: Springer-Verlag, 2006, pages 294–308. DOI: 10.1007/11693024_20.

[9]   Antony Courtney. "Frappé: Functional Reactive Programming in Java". In: *Practical Aspects of Declarative Languages, Third International Symposium, PADL 2001, Las Vegas, Nevada, USA, March 11-12, 2001, Proceedings*. Edited by I. V. Ramakrishnan. Volume 1990. Lecture Notes in Computer Science. Berlin, Heidelberg: Springer-Verlag, 2001, pages 29–44. DOI: 10.1007/3-540-45241-9_3.

[10]  Antony Courtney, Henrik Nilsson, and John Peterson. "The Yampa arcade". In: *Proceedings of the ACM SIGPLAN Workshop on Haskell, Haskell 2003, Uppsala, Sweden, August 28, 2003*. Edited by Johan Jeuring. New York, NY, USA: Association for Computing Machinery, 2003, pages 7–18. DOI: 10.1145/871895.871897.

[11]  Evan Czaplicki. *Controlling Time and Space: understanding the many formulations of FRP*. Strange Loop Conference Talk. 2014. https://www.youtube.com/watch?v=Agu6jipKfYw; last accessed on 2024-02-09.

[12]  Evan Czaplicki and Stephen Chong. "Asynchronous functional reactive programming for GUIs". In: *ACM SIGPLAN Conference on Programming Language Design and Implementation, PLDI '13, Seattle, WA, USA, June 16-19, 2013*. Edited by Hans-Juergen Boehm and Cormac Flanagan. New York, NY, USA: Association for Computing Machinery, 2013, pages 411–422. DOI: 10.1145/2491956.2462161.

[13]  Joeri De Koster, Tom Van Cutsem, and Wolfgang De Meuter. "43 years of actors: a taxonomy of actor models and their key properties". In: *Proceedings of the 6th International Workshop on Programming Based on Actors, Agents, and Decentralized Control, AGERE 2016, Amsterdam, The Netherlands, October 30, 2016*. Edited by Sylvan Clebsch, Travis Desell, Philipp Haller, and Alessandro Ricci. New York, NY, USA: Association for Computing Machinery, 2016, pages 31–40. DOI: 10.1145/3001886.3001890.

[14]  Joscha Drechsler, Guido Salvaneschi, Ragnar Mogk, and Mira Mezini. "Distributed REScala: an update algorithm for distributed reactive programming". In: *Proceedings of the 2014 ACM International Conference on Object Oriented Programming Systems Languages & Applications, OOPSLA 2014, part of SPLASH 2014, Portland, OR, USA, October 20-24, 2014*. Edited by Andrew P. Black and Todd D. Millstein. New York, NY, USA: Association for Computing Machinery, 2014, pages 361–376. DOI: 10.1145/2660193.2660240.

[15]  Conal Elliott and Paul Hudak. "Functional Reactive Animation". In: *Proceedings of the 1997 ACM SIGPLAN International Conference on Functional Programming (ICFP '97), Amsterdam, Netherlands, June 9-11, 1997*. Edited by Simon L. Peyton Jones, Mads Tofte, and A. Michael Berman. New York, NY, USA: Association for Computing Machinery, 1997, pages 263–273. DOI: 10.1145/258948.258973.

[16]  Matthias Felleisen, Robert Bruce Findler, Matthew Flatt, Shriram Krishnamurthi, Eli Barzilay, Jay McCarthy, and Sam Tobin-Hochstadt. "A programmable programming language". In: *Communications of the ACM* 61.3 (Feb. 2018), pages 62–71. ISSN: 0001-0782. DOI: 10.1145/3127323.

[17]  Matthew Flatt and PLT. *The Racket Reference: Definitions*. https://web.archive.org/web/20230830132742/https://docs.racket-lang.org/reference/define.html. Last accessed on 2023-08-30.

[18]  Adele Goldberg and David Robson. *Smalltalk-80: The Language and Its Implementation*. Addison-Wesley, 1983. ISBN: 978-02-0113-688-3.

[19]  Nicolas Halbwachs. "Synchronous Programming of Reactive Systems". In: *Computer Aided Verification, 10th International Conference, CAV '98, Vancouver, BC, Canada, June 28 - July 2, 1998, Proceedings*. Edited by Alan J. Hu and Moshe Y. Vardi. Volume 1427. Lecture Notes in Computer Science. Springer, 1998, pages 1–16. DOI: 10.1007/BFb0028726.

[20]  Nicolas Halbwachs, Paul Caspi, Pascal Raymond, and Daniel Pilaud. "The synchronous data flow programming language LUSTRE". In: *Proceedings of the IEEE* 79.9 (1991), pages 1305–1320. DOI: 10.1109/5.97300.

[21]  Caleb Helbling and Samuel Z. Guyer. "Juniper: a functional reactive programming language for the Arduino". In: *Proceedings of the 4th International Workshop on Functional Art, Music, Modelling, and Design, FARM@ICFP 2016, Nara, Japan, September 24, 2016*. Edited by David Janin and Michael Sperber. New York, NY, USA: Association for Computing Machinery, 2016, pages 8–16. DOI: 10.1145/2975980.2975982.

[22]  Martin Hirzel, Rodric Rabbah, Philippe Suter, Olivier Tardieu, and Mandana Vaziri. "Spreadsheets for stream processing with unbounded windows and partitions". In: *Proceedings of the 10th ACM International Conference on Distributed and Event-based Systems, DEBS '16, Irvine, CA, USA, June 20 - 24, 2016*. Edited by Avigdor Gal, Matthias Weidlich, Vana Kalogeraki, and Nalini Venkasubramanian. New York, NY, USA: Association for Computing Machinery, 2016, pages 49–60. DOI: 10.1145/2933267.2933607.

[23]  Paul Hudak, Antony Courtney, Henrik Nilsson, and John Peterson. "Arrows, Robots, and Functional Reactive Programming". In: *Advanced Functional Programming, 4th International School, AFP 2002, Oxford, UK, August 19-24, 2002, Revised Lectures*. Edited by Johan Jeuring and Simon L. Peyton Jones. Volume 2638. Lecture Notes in Computer Science. Berlin, Heidelberg: Springer-Verlag, 2002, pages 159–187. DOI: 10.1007/978-3-540-44833-4_6.

[24]  Daniel Ignatoff, Gregory H. Cooper, and Shriram Krishnamurthi. "Crossing State Lines: Adapting Object-Oriented Frameworks to Functional Reactive Languages". In: *Functional and Logic Programming, 8th International Symposium, FLOPS 2006, Fuji-Susono, Japan, April 24-26, 2006, Proceedings*. Edited by Masami Hagiya and Philip Wadler. Volume 3945. Lecture Notes in Computer Science. Berlin, Heidelberg: Springer-Verlag, 2006, pages 259–276. DOI: 10.1007/11737414_18.

[25]  Richard Kelsey, William D. Clinger, and Jonathan Rees. "Revised[5] Report on the Algorithmic Language Scheme". In: *ACM SIGPLAN Notices* 33.9 (1998), pages 26–76. DOI: 10.1145/290229.290234.

[26]   Jeffrey C. Lagarias. "The 3x + 1 Problem and Its Generalizations". In: *The American Mathematical Monthly* 92.1 (1985), pages 3–23. ISSN: 1930-0972. JSTOR: 2322189.

[27]   Henrik Larsen, Gijs van der Hoorn, and Andrzej Wąsowski. "Reactive Programming of Robots with RxROS". In: *Robot Operating System (ROS): The Complete Reference (Volume 6)*. Edited by Anis Koubaa. Cham: Springer International Publishing, 2021, pages 55–83. ISBN: 978-3-030-75472-3. DOI: 10.1007/978-3-030-75472-3_2.

[28]   Ton Chanh Le, Timos Antonopoulos, Parisa Fathololumi, Eric Koskinen, and ThanhVu Nguyen. "DynamiTe: dynamic termination and non-termination proofs". In: *Proceedings of the ACM on Programming Languages* 4.OOPSLA (2020), 189:1–189:30. DOI: 10.1145/3428257.

[29]   Steven Macenski, Tully Foote, Brian P. Gerkey, Chris Lalancette, and William Woodall. "Robot Operating System 2: Design, architecture, and uses in the wild". In: *Science Robotics* 7.66 (2022). DOI: 10.1126/SCIROBOTICS.ABM6074.

[30]   Ingo Maier and Martin Odersky. *Deprecating the Observer Pattern with Scala.React*. Technical report EPFL-REPORT-176887. EPFL IC IINFCOM LAMP, Station 14, 1015 Lausanne: École Polytechnique Fédérale de Lausanne, 2012. URL: https://web.archive.org/web/20230509125331/https://core.ac.uk/download/pdf/147982753.pdf. Last accessed on 2023-05-09.

[31]   Geoffrey Mainland, Greg Morrisett, and Matt Welsh. "Flask: staged functional programming for sensor networks". In: *Proceeding of the 13th ACM SIGPLAN international conference on Functional programming, ICFP 2008, Victoria, BC, Canada, September 20-28, 2008*. Edited by James Hook and Peter Thiemann. New York, NY, USA: Association for Computing Machinery, 2008, pages 335–346. DOI: 10.1145/1411204.1411251.

[32]   John McCarthy. "Recursive Functions of Symbolic Expressions and Their Computation by Machine, Part I". In: *Communications of the ACM* 3.4 (1960), pages 184–195. DOI: 10.1145/367177.367199.

[33]   Leo A. Meyerovich, Arjun Guha, Jacob P. Baskin, Gregory H. Cooper, Michael Greenberg, Aleks Bromfield, and Shriram Krishnamurthi. "Flapjax: a programming language for Ajax applications". In: *Proceedings of the 24th Annual ACM SIGPLAN Conference on Object-Oriented Programming, Systems, Languages, and Applications, OOPSLA 2009, October 25-29, 2009, Orlando, Florida, USA*. Edited by Shail Arora and Gary T. Leavens. New York, NY, USA: Association for Computing Machinery, 2009, pages 1–20. DOI: 10.1145/1640089.1640091.

[34]   Tom E. Murphy. "A livecoding semantics for functional reactive programming". In: *Proceedings of the 4th International Workshop on Functional Art, Music, Modelling, and Design, FARM@ICFP 2016, Nara, Japan, September 24, 2016*. Edited by David Janin and Michael Sperber. ACM, 2016, pages 48–53. DOI: 10.1145/2975980.2975986.

[35]    Florian Myter, Christophe Scholliers, and Wolfgang De Meuter. "Many spiders make a better web: a unified web-based actor framework". In: *Proceedings of the 6th International Workshop on Programming Based on Actors, Agents, and Decentralized Control, AGERE 2016, Amsterdam, The Netherlands, October 30, 2016*. Edited by Sylvan Clebsch, Travis Desell, Philipp Haller, and Alessandro Ricci. New York, NY, USA: Association for Computing Machinery, 2016, pages 51–60. DOI: 10.1145/3001886.3001892.

[36]    Phúc C. Nguyễn, Thomas Gilray, Sam Tobin-Hochstadt, and David Van Horn. "Size-change termination as a contract: dynamically and statically enforcing termination for higher-order programs". In: *Proceedings of the 40th ACM SIGPLAN Conference on Programming Language Design and Implementation, PLDI 2019, Phoenix, AZ, USA, June 22-26, 2019*. Edited by Kathryn S. McKinley and Kathleen Fisher. New York, NY, USA: Association for Computing Machinery, 2019, pages 845–859. DOI: 10.1145/3314221.3314643.

[37]    Henrik Nilsson, Antony Courtney, and John Peterson. "Functional reactive programming, continued". In: *Proceedings of the 2002 ACM SIGPLAN Workshop on Haskell, Haskell 2002, Pittsburgh, Pennsylvania, USA, October 3, 2002*. Edited by Manuel M. T. Chakravarty. New York, NY, USA: Association for Computing Machinery, 2002, pages 51–64. DOI: 10.1145/581690.581695.

[38]    Bjarno Oeyen, Sam Van den Vonder, and Wolfgang De Meuter. "Reactive sorting networks". In: *REBLS 2020: Proceedings of the 7th ACM SIGPLAN International Workshop on Reactive and Event-Based Languages and Systems, Virtual Event, USA, November 16, 2020*. New York, NY, USA: Association for Computing Machinery, 2020, pages 38–50. DOI: 10.1145/3427763.3428316.

[39]    Sean Parent. "A Possible Future for Software Development". BoostCon 2007 keynote. May 2007. URL: http://web.archive.org/web/20210624121558/https://stlab.cc/legacy/figures/Boostcon_possible_future.pdf. Last accessed on 2021-06-24.

[40]    Ivan Perez, Manuel Bärenz, and Henrik Nilsson. "Functional reactive programming, refactored". In: *Proceedings of the 9th International Symposium on Haskell, Haskell 2016, Nara, Japan, September 22-23, 2016*. Edited by Geoffrey Mainland. New York, NY, USA: Association for Computing Machinery, 2016, pages 33–44. DOI: 10.1145/2976002.2976010.

[41]    Benjamin C. Pierce. *Types and programming languages*. MIT Press, 2002. ISBN: 978-0-262-16209-8.

[42]    Gordon D. Plotkin. "Call-by-Name, Call-by-Value and the lambda-Calculus". In: *Theoretical Computer Science* 1.2 (1975), pages 125–159. DOI: 10.1016/0304-3975(75)90017-1.

[43]    Bob Reynders, Frank Piessens, and Dominique Devriese. "Gavial: Programming the web with multi-tier FRP". In: *The Art, Science, and Engineering of Programming* 4.3 (2020), 6:1–6:32. ISSN: 2473-7321. DOI: 10.22152/programming-journal.org/2020/4/6.

[44]   Guido Salvaneschi, Gerold Hintz, and Mira Mezini. "REScala: bridging be-
       tween object-oriented and functional style in reactive applications". In: *13th In-
       ternational Conference on Modularity, MODULARITY '14, Lugano, Switzerland,
       April 22-26, 2014*. Edited by Walter Binder, Erik Ernst, Achille Peternier, and
       Robert Hirschfeld. New York, NY, USA: Association for Computing Machinery,
       2014, pages 25–36. DOI: 10.1145/2577080.2577083.

[45]   Guido Salvaneschi, Sebastian Proksch, Sven Amann, Sarah Nadi, and Mira
       Mezini. "On the Positive Effect of Reactive Programming on Software Compre-
       hension: An Empirical Study". In: *IEEE Transactions on Software Engineering*
       43.12 (2017), pages 1125–1143. DOI: 10.1109/TSE.2017.2655524.

[46]   Kensuke Sawada and Takuo Watanabe. "Emfrp: a functional reactive program-
       ming language for small-scale embedded systems". In: *Companion Proceedings
       of the 15th International Conference on Modularity, Málaga, Spain, March 14 -
       18, 2016*. Edited by Lidia Fuentes, Don S. Batory, and Krzysztof Czarnecki. New
       York, NY, USA: Association for Computing Machinery, 2016, pages 36–44. DOI:
       10.1145/2892664.2892670.

[47]   Neil Sculthorpe and Henrik Nilsson. "Safe functional reactive programming
       through dependent types". In: *Proceeding of the 14th ACM SIGPLAN interna-
       tional conference on Functional programming, ICFP 2009, Edinburgh, Scotland,
       UK, August 31 - September 2, 2009*. Edited by Graham Hutton and Andrew P.
       Tolmach. ACM, 2009, pages 23–34. DOI: 10.1145/1596550.1596558.

[48]   Kazuhiro Shibanai and Takuo Watanabe. "Distributed functional reactive pro-
       gramming on actor-based runtime". In: *Proceedings of the 8th ACM SIGPLAN
       International Workshop on Programming Based on Actors, Agents, and Decen-
       tralized Control, AGERE!@SPLASH 2018, Boston, MA, USA, November 5, 2018*.
       Edited by Joeri De Koster, Federico Bergenti, and Juliana Franco. New York,
       NY, USA: Association for Computing Machinery, 2018, pages 13–22. DOI: 10.
       1145/3281366.3281370.

[49]   Michael Sperber. "Developing a Stage Lighting System from Scratch". In: *Pro-
       ceedings of the Sixth ACM SIGPLAN International Conference on Functional Pro-
       gramming (ICFP '01), Firenze (Florence), Italy, September 3-5, 2001*. Edited by
       Benjamin C. Pierce. New York, NY, USA: Association for Computing Machin-
       ery, 2001, pages 122–133. DOI: 10.1145/507635.507652.

[50]   David M. Ungar and Randall B. Smith. "Self". In: *Proceedings of the Third
       ACM SIGPLAN History of Programming Languages Conference (HOPL-III), San
       Diego, California, USA, 9-10 June 2007*. Edited by Barbara G. Ryder and Brent
       Hailpern. ACM, 2007, pages 1–50. DOI: 10.1145/1238844.1238853.

[51]   Sam Van den Vonder. "On the Coexistence of Reactive Code and Imperative
       Code in Distributed Applications: A Language Design Approach". PhD thesis.
       Vrije Universiteit Brussel, Brussels: Vrije Universiteit Brussel, June 2022. ISBN:
       978-94-6444-328-8. https://web.archive.org/web/20230509150202/https://cris.
       vub.be/ws/portalfiles/portal/86362735/Sam_Van_den_Vonder_PhD_thesis.pdf,
       last accessed on 2023-05-09.

[52] Sam Van den Vonder, Thierry Renaux, and Wolfgang De Meuter. "Topology-Level Reactivity in Distributed Reactive Programs: Reactive Acquaintance Management using Flocks". In: *The Art, Science, and Engineering of Programming* 6.3 (2022), 14:1–14:36. DOI: 10.22152/programming-journal.org/2022/6/14.

[53] Sam Van den Vonder, Thierry Renaux, Bjarno Oeyen, Joeri De Koster, and Wolfgang De Meuter. "Tackling the Awkward Squad for Reactive Programming: The Actor-Reactor Model". In: *34th European Conference on Object-Oriented Programming, ECOOP 2020, November 15-17, 2020, Berlin, Germany (Virtual Conference)*. Edited by Robert Hirschfeld and Tobias Pape. Volume 166. LIPIcs Leibniz International Proceedings in Informatics. Wadern, Germany: Schloss Dagstuhl - Leibniz-Zentrum für Informatik, 2020, 19:1–19:29. DOI: 10.4230/LIPIcs.ECOOP.2020.19.

[54] Mandana Vaziri, Olivier Tardieu, Rodric Rabbah, Philippe Suter, and Martin Hirzel. "Stream Processing with a Spreadsheet". In: *ECOOP 2014 - Object-Oriented Programming - 28th European Conference, Uppsala, Sweden, July 28 - August 1, 2014. Proceedings*. Edited by Richard E. Jones. Volume 8586. Lecture Notes in Computer Science. Berlin, Heidelberg: Springer-Verlag, 2014, pages 360–384. DOI: 10.1007/978-3-662-44202-9_15.

[55] Andreas Voellmy and Paul Hudak. "Nettle: Taking the Sting Out of Programming Network Routers". In: *Practical Aspects of Declarative Languages - 13th International Symposium, PADL 2011, Austin, TX, USA, January 24-25, 2011. Proceedings*. Edited by Ricardo Rocha and John Launchbury. Volume 6539. Lecture Notes in Computer Science. Berlin, Heidelberg: Springer-Verlag, 2011, pages 235–249. DOI: 10.1007/978-3-642-18378-2_19.

[56] Philip Wadler, Wen Kokke, and Jeremy G. Siek. *Programming Language Foundations in Agda*. Aug. 2022. URL: https://plfa.inf.ed.ac.uk/22.08/. Last accessed on 2024-02-09.

[57] Zhanyong Wan, Walid Taha, and Paul Hudak. "Event-Driven FRP". In: *Practical Aspects of Declarative Languages, 4th International Symposium, PADL 2002, Portland, OR, USA, January 19-20, 2002, Proceedings*. Edited by Shriram Krishnamurthi and C. R. Ramakrishnan. Volume 2257. Lecture Notes in Computer Science. Berlin Heidelberg: Springer-Verlag, 2002, pages 155–172. DOI: 10.1007/3-540-45587-6_11.

[58] Zhanyong Wan, Walid Taha, and Paul Hudak. "Real-Time FRP". In: *Proceedings of the Sixth ACM SIGPLAN International Conference on Functional Programming (ICFP '01), Firenze (Florence), Italy, September 3-5, 2001*. Edited by Benjamin C. Pierce. New York, NY, USA: Association for Computing Machinery, 2001, pages 146–156. DOI: 10.1145/507635.507654.

## About the authors

**Bjarno Oeyen** is a PhD candidate at the Software Languages Lab, Vrije Universiteit Brussel in Belgium. His main research area is reactive programming and distributed systems. Contact him at bjarno.oeyen@vub.be.

https://orcid.org/0000-0002-2100-4559

**Joeri De Koster** is an assistant professor in programming languages and runtimes. His current research is mainly focused on the design, formalisation and implementation of parallel and distributed programming languages. Contact him at joeri.de.koster@vub.be.

https://orcid.org/0000-0002-2932-8208

**Wolfgang De Meuter** is a professor in programming languages and programming tools. His current research is mainly situated in the field of distributed programming, concurrent programming, reactive programming and big data processing. His research methodology varies from more theoretical approaches (e.g., type systems) to building practical frameworks and tools (e.g., crowdsourcing systems). Contact him at wolfgang.de.meuter@vub.be.

https://orcid.org/0000-0002-5229-5627